\documentclass[pra,twocolumn]{revtex4}

\def\>{\rangle}
\def\<{\langle}

\def\be{\begin{equation}}
\def\ee{\end{equation}}
\def\bea{\begin{eqnarray}}
\def\eea{\end{eqnarray}}

\begin{document}
\title{Creating superpositions that correspond to efficiently integrable probability distributions}

\author{Lov Grover}
\author{Terry Rudolph}
\email{rudolpht@bell-labs.com}

\affiliation{Bell Labs, 600-700 Mountain Ave., Murray Hill, NJ 07974, U.S.A.}

\date{\today}

\begin{abstract}
We give a simple and efficient process for generating a quantum
superposition of states which form a discrete approximation of
any efficiently integrable (such as log concave) probability
density functions.
\end{abstract}


\maketitle

It is well known that probabilistic algorithms are sometimes more
powerful (and often more practical!) than their deterministic
counterparts in classical computing. Such algorithms generically
make use of some source of random bits. It has been shown
\cite{aharonov} however, that giving a quantum computer access to
random bits does not, in fact, further empower it.

There has been some attention recently to the problem of using
quantum computers to generate certain probability distributions,
in particular by looking at the properties of quantum random
walks \cite{aharonov2}.

In this short note, we will look at the question of whether,
given a certain probability distribution $\{p_i\}$, we can
efficiently create a quantum superposition of the form
\be
|\psi(\{p_i\})\>=\sum_i\sqrt{p_i}|i\>,
\ee
where the $|i\>$ are an orthonormal set of states. We are, of
course, particularly interested in the case where the index $i$
ranges over some exponentially large set of possibilities $N$.

We have not solved the general problem of when a state of the form
(1) can be efficiently created. However we will show here that if
an efficient classical algorithm exists to \emph{integrate} a
certain probability distribution $p(x)$, then we can efficiently
create a state of the form (1) for the discretized version
$\{p_i\}$ of
$p(x)$. A well known set of probability density functions
which are efficiently integrable by
monte carlo methods are \emph{log-concave} distributions. (A
log-concave distribution is one for which
$\frac{\partial^2\log p(x)}{\partial x^2}<0$.) Most important statistical distributions
(including the exponential and normal families) are log-concave.
The task of sampling and/or integrating such distributions
frequently arises for problems of determining the
(multidimensional) volumes of solids
\cite{applegate}.

Clearly if an efficient classical algorithm exists, then we can
certainly create a quantum computer in the mixed state
$\rho=\sum_ip_i|i\>\<i|$. Our interest in creating a state of the
form (1) is thus not simply to be able to measure the state
$|i\>$ with probability $p_i$. Rather we hope that generating the
distribution in a coherent fashion will allow \emph{further}
(uniquely quantum) processing of this distribution. At the end we
discuss some possibilities along these lines.

We will illustrate the procedure considering only a probability
distribution over a single random variable, although the method
can be easily extended to multivariable distributions.


We let $n=\log N$, where $N$ is the total number of points over
which we wish to discretize this distribution. Imagine the
distribution is divided into some number
$2^m$ of regions, and that we already have a $m$ qubit state
\be
|\psi_m\>=\sum_{i=0}^{2^m-1} \sqrt{p_i^{(m)}}|i\>,
\ee
where, as usual, the integer $i$, is interpreted in binary as the
tensor product sequence of qubit states. Here $p_i^{(m)}$ is the
probability for the random variable
$x$ to lie region $i$; thus $p_0^{(m)}$ is the
probability for
$x$ to lie in the far leftmost region, $p_1^{(m)}$ is the
probability for it to lie in the region adjacent to this and so
on. Our goal is to show that we can now efficiently subdivide
these $2^m$ regions to yield a $2^{m+1}$ region discretization of
$p(x)$.  That is we wish to add one qubit to the state (2), such
that we achieve the evolution
\be
\sqrt{p_i^{(m)}}|i\>\rightarrow
\sqrt{\alpha_i}|i\>|0\>+\sqrt{\beta_i}|i\>|1\>,
\ee
where $\alpha$ (resp. $\beta$) is the probability for $x$ to lie
in the
\emph{left} (resp. \emph{right}) half of region $i$. If we can achieve
such an evolution, then the new state of our $m+1$ qubits is
\be
|\psi_{m+1}\>=\sum_{i=0}^{2^{m+1}-1} \sqrt{p_i^{(m+1)}}|i\>.
\ee
This process is then repeated until $m=n$, i.e. until we have
created the desired superposition (1) over all $N=2^n$ states. If
such an evolution is possible, then clearly it is efficient.

To show how an evolution of the form (3) can be achieved, we
first define $x_L^i$ and $x_R^i$ to be the left and right
boundaries of region $i$. We then define the function
\[
f(i)=\frac{\int_{x_L^i}^{\frac{x_R^i-x_L^i}{2}}p(x)dx}{\int_{x_L^i}^{x_R^i}p(x)dx}.
\]
$f(i)$ is simply the probability that, given $x$ lies in region $i$, it also
lies in the left half of this region. Since
$\int_a^b p(x) dx$ is efficiently computable classically, we can
take an ancilla register initially in the state $|0\ldots 0\>$,
and construct a circuit which efficiently performs the computation
\be
\sqrt{p_i^{(m)}}|i\>|0\ldots
0\>\rightarrow\sqrt{p_i^{(m)}}|i\>|\theta_i\>,
\ee
where $\theta_i\equiv\arccos\sqrt{f(i)}$. We now perform a
controlled rotation of angle $\theta_i$ on the $m+1$'th qubit:
\be
\sqrt{p_i^{(m)}}|i\>|\theta_i\>|0\>\rightarrow\sqrt{p_i^{(m)}}|i\>|\theta_i\>(\cos\theta_i|0\>+\sin\theta_i|1\>)
\ee
and we then uncompute the register containing $|\theta_i\>$ to
leave us in a state of the form (3) as desired. Note that the
efficient classical algorithm to perform the integration may well
be probabilistic, however it can always be implemented on a
quantum computer in the required coherent manner by using an
ancilla register initially prepared containing the string of
random bits to be used in the computation.

Having seen that we can create a state of the form (1), we now
give some examples of why creating such a state might be
interesting.

(i) \emph{Non-uniform priors for quantum searching} The quantum
search algorithm \cite{lov} is typically envisioned as a search
over a large number of possible solutions to a problem, where
every potential solution is a-priori equally likely. In many
instances however, we might believe that the correct solution is
more likely to be found in a certain region of state space than
another. Thus we might wish to input a non-uniform prior
distribution into the search algorithm. If this prior
distribution is log-concave (e.g. gaussian, poissonian etc) then
the procedure outlined above can be used to generate it.

(ii) \emph{Sampling of non-log-concave distributions} A state of
the form (1) is clearly of little value if we were to just
measure the qubits and obtain outcome $|i\>$ with probability
$p_i$, since such sampling is possible classically. The
fundamental difference between a quantum superposition and a
classical sample however, is \emph{quantum interference}.
 In particular, we may perform unitary transformations on this
state to take it to some other state of the form
\be
|\psi'\>=\sum_j \sqrt{q_j(\{p_i\})} |j\>.
\ee
Each of the new probabilities $q_j$ can in some sense be functions
of the full set of $p_i$'s, and we may obtain ``interference''
between the probabilities that has no classical analogue. For
example, a simple Walsh-Hadamard transformation on each of the
the qubits produces the new distribution
\be
q_j=\frac{1}{N}(\sum_i (-1)^{i\cdot j} \sqrt{p_i})^2
\ee
and this distribution is \emph{not} log-concave.

(iii) \emph{Estimating the magnitude of a fourier component.} A
state of the form (1) can be fourier transformed efficiently
\cite{shor}. If we are interested in the magnitude of
a partciular fourier component, then techniques exist for the
process of \emph{amplitude estimation}, and these proceed
quadratically faster than their classical counterparts.

\begin{acknowledgments}
 This research was supported by the NSA \& ARO under contract No.
DAAG55-98-C-0040.
\end{acknowledgments}

\end{document}